\documentclass[twocolumn,showpacs,10pt,prb,floatfix,amssymb]{revtex4}
\usepackage{graphicx}

\begin{document}

\newcommand{\nx}{\textrm}

\title{Tunable Graphene System with Two Decoupled Monolayers}

\author{H. Schmidt$^1$}
\author{T. L\"udtke$^1$}
\author{P. Barthold$^1$}
\author{E. McCann$^2$}
\author{V. I. Falko$^2$}
\author{R. J. Haug$^1$}
\affiliation{$^1$Institut f\"ur Festk\"orperphysik, Leibniz
Universit\"at Hannover, Appelstr. 2, 30167 Hannover, Germany\\
$^2$Department of Physics, Lancaster University, Lancaster LA1
4YB, United Kingdom}
\date{\today}
\begin{abstract}The use of two truly two-dimensional gapless semiconductors, monolayer and bilayer graphene, as current-carrying components in field-effect transistors (FET) gives access to new types of nanoelectronic devices. Here, we report on the development of graphene-based FETs containing two decoupled graphene monolayers manufactured from a single one folded during the exfoliation process. The transport characteristics of these newly-developed devices differ markedly from those manufactured from a single-crystal bilayer. By analyzing Shubnikov-de Haas oscillations, we demonstrate the possibility to independently control the carrier densities in both layers using top and bottom gates, despite there being only a nano-meter scale separation between them.

\end{abstract}

\maketitle

The development of the micromechanical cleavage technique for
manufacturing ultra-thin graphitic films has made it possible to
produce FETs based upon monolayer graphene and to study their
electronic properties \cite{Novoselov2004,Novoselov2005,kim_nature,Rise}
indicating that graphene is a gapless
semiconductor with a peculiar Dirac-type spectrum of charge
carriers. Using micromechanical exfoliation, multilayers of
graphene are also often produced. Two layers of graphene prepared
by exfoliation usually exhibit crystalline ordering with a
characteristic AB stacking \cite{Novoselov2006,bilayer_mccann},
later referred to as single-crystal (SC) bilayer. This should be
contrasted to pairs/stacks of individual misoriented graphene
flakes identified among some of the peeled graphitic films using
Raman spectroscopy \cite{raman_misoriented} and in multilayers
grown epitaxially on SiC \cite{multi_single_epitax_haas}. While
the individual carrier density in such layers is usually beyond
experimental control, in this paper we report on the realisation
of two-layer graphene-based FETs where the density on the two
monolayers can be varied separately, despite them lying only a
dozen Angstroms
apart.\\

The nanostructures presented in this letter were prepared by
peeling off pieces from natural bulk graphite \cite{graphit} with
adhesive tape and placing them on a silicon wafer covered with
SiO$_2$. Since, at the intermediate stage of the utilized
deposition process only some part of the graphene flake touches
the SiO$_2$ surface, the flake flips over during the removal of
the adhesive tape, thus producing two misoriented graphene layers
lying on top of each other and separated by occasional surface
deposits (Fig. 1a sketches the top-view of such a structure).

\begin{figure}
\includegraphics[scale=1]{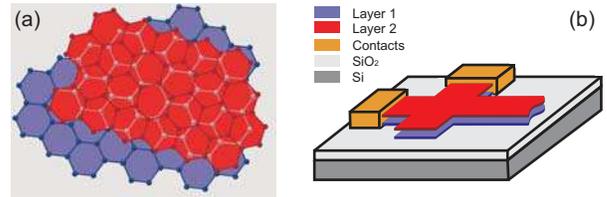}
\caption{\label{fig:bild1}a) Top view of a pair of arbitrarily
oriented graphene monolayers (Layer 2, red and Layer 1, blue) in a
two-layer stack. b) A two-layer stack on top of a Si/SiO$_2$
(grey/light grey) wafer, etched and contacted~(yellow) for
transport measurements.}
\end{figure}

Then, this material is processed into Hall-bar samples using
plasma etching and contacted (to both layers) by evaporating
chromium and gold electrodes (Fig. 1b). Figure 2a shows two
optical images of such a sample before and after plasma etching. A
darker than other areas top-left edge  indicates the position of
the fold. Scanning the structure with an atomic force microscope
(AFM) reveals a double step at the etched edges (Fig. 2c and d)
with a first-step height of 1.1 nm. Although layers of graphene
should have a height of 0.34 nm, previous AFM measurements showed
values up to 1 nm for single layer graphene lying on a substrate
\cite{Novoselov2004} attributed to a water layer underneath
graphene and/or the rippling graphene surface.

 From these data and also from the transport measurements shown below we conclude that
the first step is due to a monolayer of graphene. The second step
has a height of 0.6 nm which is larger than the thickness of one
layer but less than the height of two AB stacked layers. This
indicates that the second layer is also a single layer which is
separated further from the first layer than in a conventional
bilayer.\\

To distinguish the designed device from monolayers or SC bilayers
we performed transport measurements on the sample shown in Fig. 2,
applying magnetic fields up to 13~T at temperatures down to 1.5~K.
To reduce the influence of contact resistance on the measurements,
a multi-terminal device has been used, and different combinations
of two and three point measurements were performed at a broad
range of back-gate voltage, $-70$~V$<V_{BG}<70$~V, applied between
the substrate of n-Si and graphene. As usual, positive back-gate
voltage induces electrons, negative holes. Figure 2b shows the
measured field effect on graphene resistivity demonstrating the
operation of the device as an FET \cite{Novoselov2004}. A
characteristic peak in the observed resistivity points at the
approximate neutrality condition for the graphene layers, which is
shifted to a finite back-gate voltage $V_{BG}^0=11.5$~V indicating
natural doping of the graphene flakes. The measured maximum
resistivity, $\rho_{max}\approx 3.3$~k$\Omega$ at 1.5~K, is about
half of the earlier-reported typical values for monolayers
\cite{Novoselov2005,Rise}, which is in line with the assumption
that the device consists of two monolayers conducting in
parallel.
\begin{figure}[t]
\includegraphics[scale=1]{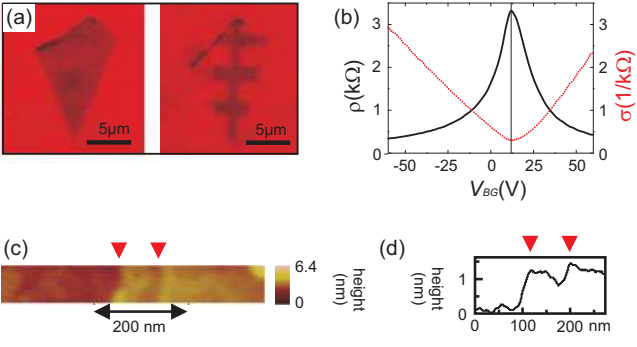}
\caption{\label{fig:bild2} a) Optical image of the two-layer
sample before and after etching. b) The resistivity (black) and
the conductivity (red, dots) plotted versus the back-gate voltage
applied between silicon substrate and graphene. c) The AFM image
of the edge showing two steps (red arrows) and d) its averaged
height profile.}
\end{figure}

Most importantly, the layer structure of ultrathin graphitic films
can be characterized by analyzing the Shubnikov-de Haas (SdH)
oscillations in the flake resistance $R(B)$ studied as a function
of a magnetic field $B$ applied perpendicular to the sample. The
Berry's phase $\pi$ characteristic of electrons in monolayer
graphene is directly related to the appearance of SdH oscillations
minima at filling factors $\nu_{min}=4(i+1/2)$ \cite{Novoselov2005,kim_nature}, in contrast to SC bilayer
graphene \cite{bilayer_mccann} where the electron's Berry's phase
is $2\pi$ and the SdH oscillations minima appear at filling
factors $\nu_{min}=4i$, with $\nu =nh/eB$ being the filling
factor, $n$ the carrier concentration, and $i$ an integer. When
studied at a fixed back-gate voltage (e.g. $V_{BG}=-70$~V and
$V_{BG}=-40$~V, Fig 3a,b), the magnetoresistance measured in the
two-layer device and plotted versus $1/B$ displays oscillations
with two very different periods (compared in detail in the
top/bottom panels of Fig. 3 a,b) manifesting the co-existence of
two very different carrier densities, $n_1$ and $n_2$ in the
sample. These two density values can be obtained from the period
$\Delta B^{-1}$ of the SdH oscillations, as $n=4e /h \Delta
B^{-1}$ (where we take into account both the spin and valley
degeneracy in the graphene band structure). Their experimentally
determined gate-voltage dependence is shown in Fig. 3d.
Systematically repeated at all the studied voltage values, the
periodicity of both fast (lower panel) and slow (upper panel)
$1/B$-oscillations in the two-layer structure was such that the
resistivity minima could only be attributed to the sequence of
filling factors $\nu_{min} =4(i+1/2)$, which is typical for
monolayer graphene.

To highlight this behavior of the presented
data, in the top/bottom panels of Fig 3a,b the $1/B$ intervals
with the rising parts of the corresponding oscillations are
colored in red/blue.  For comparison, Fig. 3c shows the results of
a similar study performed for a SC bilayer, where the sequencing
of the SdH oscillations minima coincides with the earlier observed
\cite{Novoselov2006} appearance of maxima at filling factors
$\nu_{min}=4i$ specific for such a material. Finally, the observed
carrier densities $n_1$ and $n_2$ in the two-layer device and
their dependence on back-gate voltage presented in Fig. 3d can be
compared to theoretically calculated values plotted on the same
figure as black lines. The presented calculation takes into
account the electrostatics of the device containing two monolayers
at distance $d\approx 1.5$~nm from each other lying on top of the
SiO$_2$ dielectric layer (with permittivity $\epsilon_b = 3.9$ and
thickness $L_{BG}$), Fermi-energy dependence of the
compressibility of the electron gas with the Dirac spectrum
$E(p)=vp$, and additional doping charge, $\delta n \approx
2\cdot10^{11}$~cm$^{-2}$ due to deposits left on top of the upper
layer in the course of the manufacturing process.
\begin{figure}[t]
\includegraphics[scale=1]{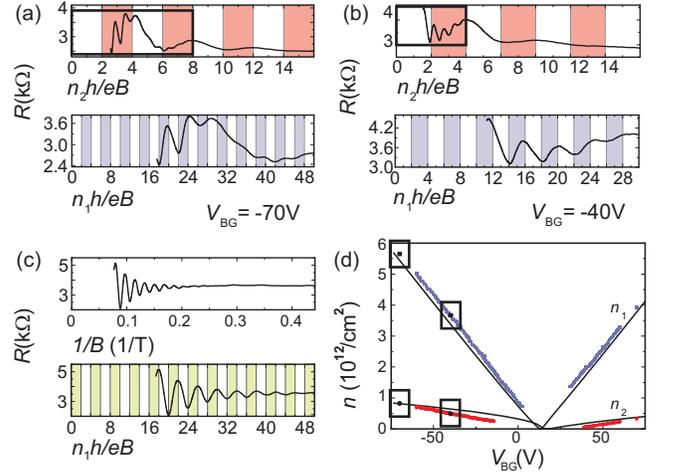}
\caption{
 a) Longitudinal resistance measured as a function of a
perpendicular magnetic field for $V_{BG}=-70$~V. The upper figure
shows the oscillation corresponding to the lower carrier
concentration $n_2$; the zoomed part in the bottom panel shows the
oscillations corresponding to the high concentration, $n_1$. b)
Same plots for $V_{BG}=-40$~V. c) Comparable measurement on a SC
bilayer showing only one sequence of oscillations with typical
bilayer behavior \cite{bilayer_mccann,Novoselov2006}.  d) Two
carrier concentrations in the two-layer (folded) graphene sample
extracted from the period of the SdH oscillations. The boxes mark
the points obtained using the data shown in panels a and b. The
black lines are the results of theoretical modelling of the device
as described in the text. A shift in the gate voltage of 11.5~V
takes into account doping of graphene by deposits.}
\end{figure}
\newpage
The parameters used in this calculation were obtained by fitting the observed
$V_{BG}$-dependence with the analytical result relating the two
densities on two parallel graphene flakes,
\begin{eqnarray*}
2 \sqrt{\pi} d(e^2/\epsilon_0) [n_2+\delta n]&=& hv(s_1
\sqrt{|n_1|}
- s_2 \sqrt{|n_2|})\\
n_1+n_2+\delta n&=&\frac{\epsilon_0\epsilon_b V_{BG}}{eL_{BG}}
\end{eqnarray*}
where the first equation states the equivalence of the
electro-chemical potentials on the two graphene layers with $s_i=
n_i/|n_i|$, whereas the second relates the total charge density in
the device to its electrostatic capacity.\\

\begin{figure}
\includegraphics[scale=1]{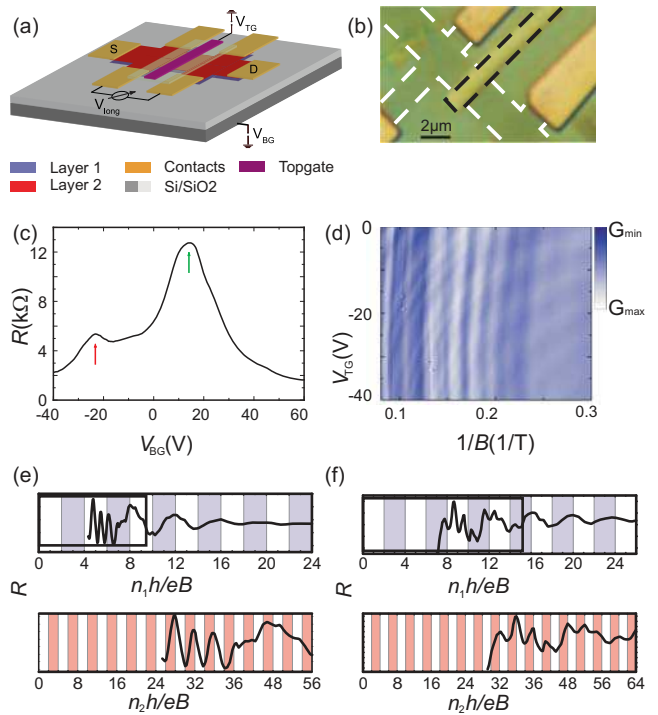}
\caption{a) Schematic drawing of the second sample used in the
experiment. The graphene layer and the top-gate metallic plate are
separated via a PMMA Layer. A current $I$ was driven through the
S-D contacts (yellow) while the longitudinal resistance ($R$) was
measured. The two layers are marked in red (Layer 2) and blue
(Layer 1). b) Image of the sample showing the etched flake, the
contacts and the top gate. The edges of the flake and of the top
gate are highlighted using dashed lines. c) Longitudinal
resistance versus back-gate voltage for $V_{TG}= 15~V$. The maxima
in the resistance correspond to the two neutrality points
underneath the top gate (red arrow) and outside the gated region
(green arrow). Using the shift with gate voltages, one obtains a
ratio of 2.5 between the capacitive couplings of the flake to the
top and back gates. d) 2D-plot showing $dR/dB$ versus the inverse
magnetic field ($B^{-1}$) and the top-gate voltage at $\nx V_{\nx
BG} = -40~V$. Different sets of Shubnikov-de Haas oscillations can
be seen. Traces for two different top-gate voltages ($V_{TG} =
-14~V$ and $V_{TG} = -31~V$) are depicted in Figs. e and f showing
maxima at $nh/eB = 0$ for both oscillations.}
\end{figure}

To tune the two carrier densities independently on the two
parallel layers, we fabricated a two-layer sample with an
additional top gate. A schematic view of the manufactured device
is shown in Fig. 4a, with the design resembling that of the
top-gated SC bilayer structures used recently \cite{Morpurgo2007}.
To fabricate the top gate, polymethylmethacrylat (PMMA) was spun
onto a Hall-bar device which was produced in a similar way as the
sample shown in Fig. 2. The PMMA layer was partially exposed to an
electron beam converting the PMMA into a layer insoluble by
acetone. Putting the sample in an acetone bath leaves an
insulating PMMA layer of about 60~nm on top of the graphene flake.
A local top gate with an area of $2.5 ~\mu$m$^2$ was fabricated on
top of this PMMA layer using standard electron beam lithography.
Figure 4b shows an image of the top-gated sample. The resistance
measured in this device as a function of the back-gate voltage for
a fixed top-gate voltage, $V_{TG}$ is shown in Fig. 4c (for
$V_{TG}=15$~V). Its back-gate voltage dependence contains two
pronounced peaks (marked by arrows) indicating the difference
between the carrier concentrations in the top-gated and free parts
of the device. The back-gate voltage interval between the two
peaks corresponds to the bipolar transistor regime, with a p-n-p
junction \cite{Falko_PhysRev} formed underneath the top gate
\cite{C_Marcus_Science,P_Kim_PRL}. The tunability of the device
using the top gate is demonstrated in Fig. 4d. Here, the back-gate
voltage was kept fixed at $V_{BG}=-40$~V and the top-gate voltage,
$V_{TG}$ was varied. Due to the large negative back-gate voltage,
the resistance maximum corresponding to the 'neutrality point' in
the top-gated region appears at positive voltage $V_{TG}=26$V (not
shown in the figure). To improve the visibility of the SdH
oscillations and their evolution with $V_{TG}$, we differentiate
the device's resistance with respect to the magnetic field value
and show the result using the blue-scale plot in Fig. 4d where one
sees several sets of oscillations. The oscillations which are
almost independent of the top-gate voltage originate from the
areas outside the top-gated region. The oscillations with a strong
$V_{TG}$ dependence characterize the layers underneath the top
gate. The deconvolution of these two contributions is achieved by
the numerical subtraction of the $V_{TG}$ independent part of the
resistance. The resulting magnetooscillations are shown in Figs.
4e,f as  function of filling factor for $V_{TG}=-14$~V and
$V_{TG}=-31$~V. Similarly to the first two-layer device, the
top-gated structure shows the superposition of two monolayer-type
SdH oscillations with different periods corresponding to densities
$n_2= 6.32\cdot 10^{12}$~cm$^{-2}$ and $n_1= 1.08 \cdot
10^{12}$~cm$^{-2}$ for $V_{TG}=-14$~V and $V_{BG}=-40$~V, and to
$n_2= 7.44 \cdot 10^{12}$~cm$^{-2}$ and $n_1= 1.80 \cdot
10^{12}$~cm$^{-2}$ for $V_{TG}=-31$~V and $V_{BG}=-40$~V. This
demonstrates not only that the different carrier densities on the
two closely laid graphene layers can be detected, but also that
those densities can be independently controlled and tuned using a
combination of top/backgates.\\
\newpage
In conclusion, we demonstrated the realization of a graphene-based
field-effect transistor containing two decoupled graphene
monolayers at a nanometer distance from each other. Using the
magnetotransport measurements, we determine the carrier densities
on the two parallel layers and show the ability to control those
densities separately using a combination of electrostatic (back \&
top) gates. When operated in a broad voltage range, such a device
could be employed in a search for the recently predicted
superfluidity \cite{Min2008} in the 'excitonic insulator' state
expected to form in a pair of graphene layers with opposite
polarity. Also, the technique of layer folding offers a promising
method for making devices with separately contacted pairs of
monolayers acting as atomically-thin optically-transparent
\cite{Abergel2007,GeimOptics} electrodes to study vertical
transport and electro-optical characteristics of nanoparticles of
various materials that can be captured between them.

The authors acknowledge financial support by the excellence cluster QUEST within the German Excellence Initiative.
V.~F. thanks the A.~von~Humboldt Foundation for hospitality. This work was
supported by ESF grant SpiCo, EPSRC-Lancaster Portfolio Partnership grant,
and EPSRC. 

\newpage
\end{document}